# A Temperature Monitoring System Incorporating an Array of Precision Wireless Thermometers

Amir Javadpour, Hamidreza Memarzadeh-Tehran and Fatemeh Saghafi
Faculty of New Sciences and Technologies
University of Tehran
Tehran, Iran
hmemar@ut.ac.ir

*Abstract*—This paper addresses the design and implementation of a real-time temperature monitoring system with applications in telemedicine. The system consists of a number of precision wireless thermometers which are conceived and realized to measure the patients' body temperature in hospitals and the intensive care units (ICUs). Each wireless thermometer incorporates an accurate semiconductor temperature sensor, a transceiver operating at 2.4 GHz and a microcontroller that controls the thermometer functionalities. An array of two thermometers are implemented and successfully evaluated in different scenarios, including free-space and *in vivo* tests. Also, an in-house developed computer software is used in order to visualize the measurements in addition to detecting rapid increase and alerting high body temperature. The agreement between the experimental data and reference temperature values is significant.

## I. Introduction

Wireless technology has been the enabling domain in reshaping conventional healthcare systems [1] in conjunction with information technology (IT) [2]. Emerging technologies such as m-health [3], ubiquitous health monitoring [4] as well as telemedicine have recently become widespread and attracted the attention of many researchers. The Continuous and realtime monitoring systems, as the key elements of modern caregiving systems, can effectively revolutionize the conventional healthcare systems [5].

Wireless body sensor networks (WBSNs) [6] have received a considerable attention as the viable alternative in achieving continuous health monitoring systems. Currently, WBSNs support a wide range of applications including fall prevention [7], wireless electrocardiography (ECG) (i.e., wireless ECG), and also remote respiration and temperature monitoring [8]. As it can be deduced, different types of vital sign sensors (e.g, blood pressure, glucose, ECG and temperature) which are wirelessly networked together can incorporate in WBSNs for specific health monitoring purposes. However, currently in hospitals, patients' vital signs are recorded and supervised several times during the course of a day by clinical staff. Human errors, lack of adequate skills, tiredness and inefficient staff in additional to shortcoming of sufficient accuracy due to wrong measurement and personal interpretation of the results can deteriorate patients' life, especially when the number of hospitalized patients exceeds.

Normal human body temperature is approximately 37 C around which all human organs, specifically the brain, can effectively function. An increase in the body temperature beyond 37 C (i.e., fever) can be the consequence of an infection or general health complication [9]. Sudden increase of body temperature can cause severe health concerns which adversely affect the normal life of patients, particularly children, temporarily or even permanently. Therefore, continuous monitoring of patients' critical and even the non-critical vital signs such as temperature seems to be necessary. As a consequence, it is found essential to design and implement a realtime monitoring system, particularly for the patients who are hospitalized in intense care units in case of emergency [10].

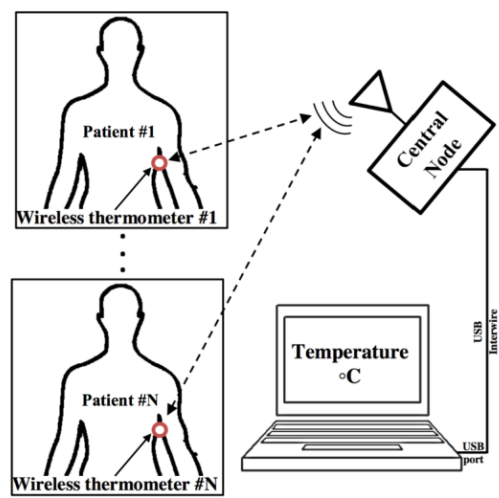

Fig. 1. Schematic depicting the developed temperature monitoring system consisting of N individual wireless thermometers.

This paper describes the design and implementation of a real-time monitoring system for obtaining the human body temperature in a continuous manner. The system comprises of a number of precision wireless thermometers and a communication infrastructure which is used for handling the measurement results. In comparison to previously reported temperature monitoring systems [11] and [12], the proposed system consists in a built-in identification mechanism in order to specify the measurement performed by a given thermometer. This identification mechanism can enable the monitoring system to potentially utilize a large number of thermometers. However, the maximum number of the thermometers are limited by the rate of temperature increase and required accuracy.

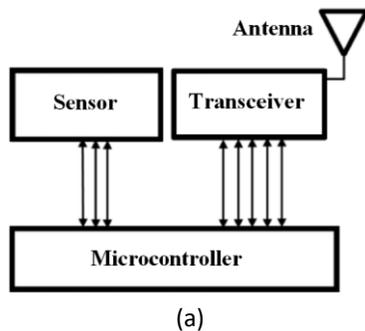

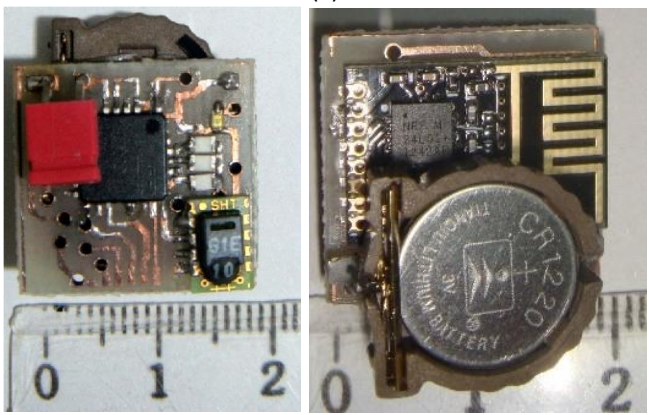

Fig. 2. The wireless thermometer, (a) schematic of the thermometer's fundamental parts. Photographs of the developed thermometer, (b) front view showing the microcontroller and the temperature sensor, (c) back view showing the transceiver and the PIFA antenna.

Additionally, the system is able to detect rapid increase and high body temperature of the patient who wearing the wireless thermometer.

## II. THE DEVELOPED TEMPERATURE MONITORING SYSTEM

Figure 1 represents the schematic of the developed monitoring system which incorporates a number of essential components namely, the wireless thermometers and the communication infrastructure. The latter also includes a central node for transmitting and receiving the signals at the radio frequency of 2.4 GHz and a low-frequency interconnents. Additionally, a PC is used to coordinate the measurements performed by the thermometers. In sections II-A and B, the design and implementation of the thermometers and communication infrastructure will be described.

### A. Wireless Thermometer

The schematic and a photograph of the developed wireless thermometer are represented in Fig. 2. This system operates at 2.4 GHz, one of the popular frequencies in the ISM (Industrial, Scientific and medical) band. Similar to a typical sensor node, the wireless thermometer consists of three main parts namely, a transceiver (nRF24, Nordic Semiconductor), a microcontroller (Atmega88, Atmel) and a digital temperature sensor (SHT11, Sensirion AG) with the accuracy and resolution of 0.4 and 0.01 C respectively. As it can be seen in Fig. 2, the transceiver uses a planar inverted-F antenna (PIFA) for transmitting and receiving the signals at the target frequency of 2.4 GHz, where the output power is approximately set at 0 dBm.

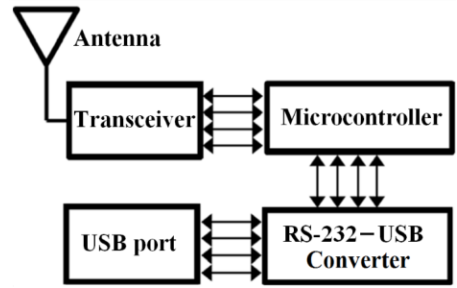

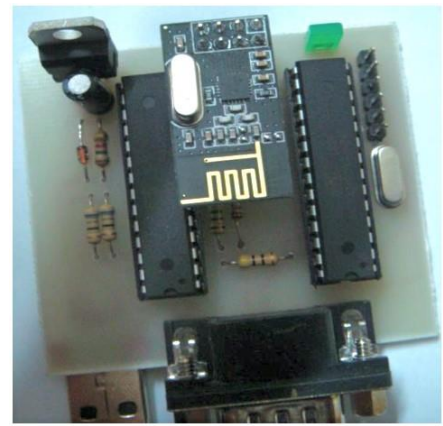

Fig. 3. The schematic and a photograph of the implemented central node.

### B. Communication Infrastructure

In order to visualize the measured temperature values for making necessary alerts, the thermometers are required to communicate with a PC or PDA (Personal Digital Assistant) through a central node as it is schematically represented in Fig. 1. The central node composes of a transceiver and a RS232 to USB port converter as the transceiver sends out the data in sets of successive sequences. Figure 3 illustrates the schematic and a photograph of the central node utilized in the monitoring system.

In general, the monitoring system can utilize a number ($N$ 2) of wireless thermometers, as shown in Fig. 1, that their unsupervised communication with the central node can result in data collision and consequently loss of information. In order to overcome the unsupervised communication, a mechanism organizing a safe access of the thermometers to the central node via the shared wireless medium is essential to be used. Also, the mechanism is required to control the duration at which an individual thermometer is allowed to transmit. Such mechanism is known as the MAC (Medium Access Control) protocols which are generally categorized in two different classes namely, the contention-based and schedulebased. In the former protocols, the thermometers need to compete for the earliest transmission chance. However, it is not a reasonable alternative for the health monitoring system thermometer estimates the power level in the channel (i.e., shared wireless medium) by using the RSSI (Received Signal Strength Indictor) mechanism before deciding to transmit. Once the medium is free, i.e., no communication in progress, the wireless thermometer transmits the data. Lastly, the central node sends an acknowledgement packet to the thermometer, confirming the successful reception of data. Figure 5 shows schematically the procedure which is followed by the pair of the central node and the measuring thermometer (Selected). It is also noteworthy to mention that this procedure enables one to identify the patient who is wearing the thermometer. Additionally, in the proposed ID-MAC the central node (master node) is responsible for requesting due to excessive delay, possible loss of critical information and also power consumption when the number of thermometers increases. Schedule-based protocols as opposed to their contention-based counterparts, can successfully be utilized in WBSNs [7]. For instance, the TDMA (Time Division Multiple Access) schedule-based protocols dedicate a guaranteed time slot to the measuring thermometer. Also, the TDMA protocols enables the system to prevent unsupervised or simultaneous communication by thermometers, causing collision [13] [14]. However, the basic TDMA protocols cannot recognize the identification of the node accessing the medium, despite it is a mandatory requirement in the developed monitoring system. Therefore, the design and implementation of a new MAC protocol equipped with an identification mechanism, contrary to most of the MAC protocols used in WBSNs, is found to be essential.

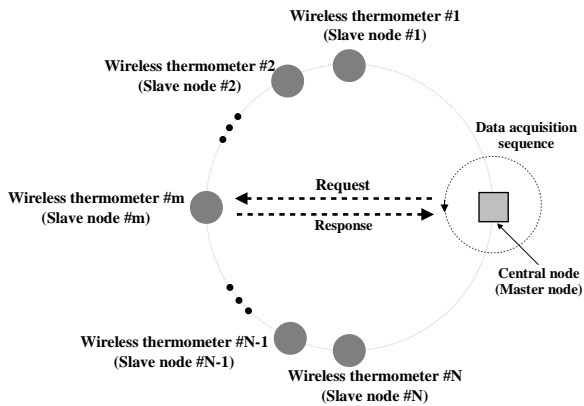

Fig. 4. Schematic depicting an array of N wireless thermometers used in the temperature monitoring system.

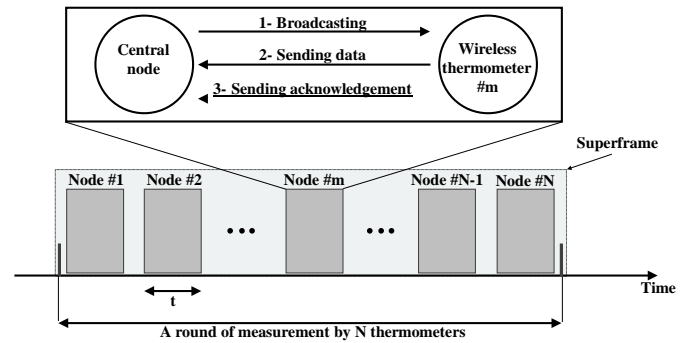

Fig. 5. Synchronization mechanism in the proposed protocol (i.e., ID-MAC).

*Proposed MAC Protocol- ID-MAC:* The proposed MAC protocol equipped with an identification mechanism (i.e., IDMAC protocol) is based on dividing the period of time which is available for a round of measurement by all the thermometers ($N$) employed in the monitoring system as shown in Fig. 4. In the proposed protocol the unique identification code, which is built-in by the manufacturer in each digital temperature sensor, is used. In a round of measurement (i.e.,central node broadcasts a packet which includes the identi-$N \rightarrow t$), the fication code of an specific node, either chosen sequentially or randomly, as shown in Fig. 5. Next, the thermometer with the matched code makes necessary arrangements to send out its data. Additionally, the the temperature value from the wireless thermometers (slave nodes) as opposed to the other protocols such as Zigbee [15] and IEEE 802.15.4 [16] in which the wireless thermometers are supposed to try to access the channel and transmit the data to the central node. Theoretically, the ID-MAC protocol is able to support a large number of thermometers in the monitoring system, however, by increasing the thermometers the delay associated with obtaining the measurement result of a thermometer increases as shown in Fig. 5. Therefore, the number of thermometers used cannot be unlimited.

III. EVALUATION OF THE TEMPERATURE MONITORING SYSTEM

In order to evaluate the performance of the monitoring system that includes an array of two wireless thermometers (Fig. 6) and also to validate the obtained results, a number of assessments partly includes the stability, response time and agility test, were performed.

*A. Stability*

One of the key characteristics of the monitoring systems is the stability of measurement results. The stability is defined as the consistency of measurement results when the system is set to measure the body temperature (A healthy person with a fixed body temperature of 37 C) over a period of time. In order to assess the stability of the developed system, the thermometer was placed on the hand of a volunteer as shown in Fig. 7. A set of measurement was performed for 60 seconds with the sampling rate of 1 second which is sufficient in monitoring non-critical vital signs such as the body temperature. For comparison purposes and also in order to identify the discrepancies in the implementation of the thermometers, the same measurement was also repeated with another thermometer. As it is illustrated in Fig. 7, the obtained results fluctuate approximately ±0.125 C around a mean value of 37 C. As a practical remedy to overcome the fluctuations associated with measurements, an average of a few samples (e.g., 5-point average) was calculated by oversampling temperature values to obtain a smoother curve. This experiment also indicates a nearly identical performance of the developed thermometers.

To investigate the sources of fluctuations, we developed a wired thermometer as it is represented in Fig. 8. The wired thermometer incorporates a SHT11 digital temperature sensor, similar to the temperature sensor used to develop the wireless thermometers. The fundamental parts of the wired thermometer are schematically shown in the inset of Fig. 8. By using the wired thermometer a healthy volunteer's body temperature was measured over a 60-second long period and the obtained results are represented in Fig. 9 in comparison with a reference temperature at 37 C (dashed line). The maximum deviation of the measured temperature values from the reference is 0.054 C, revealing that the fluctuations are not due to the performance of the SHT11 temperature sensor and also the developed wireless thermometers wheres the communication infrastructure is considered to be responsible.

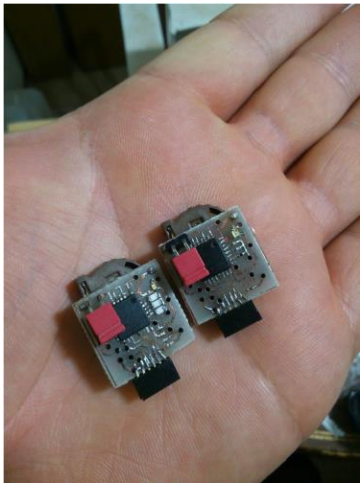

Fig. 6. Photograph illustrating the implementation of the developed array incorporating two wireless thermometers.

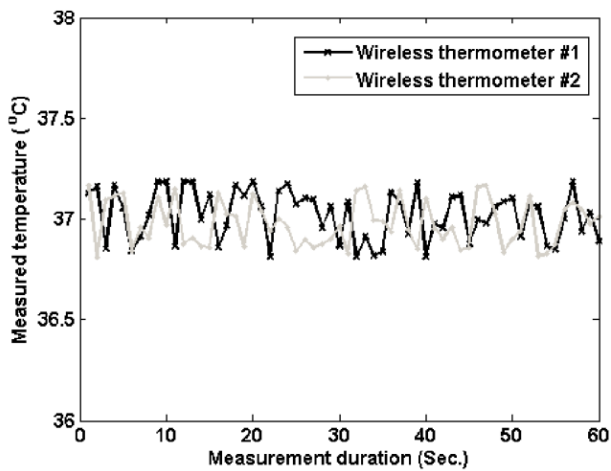

Fig. 7. Stability test results of the temperature monitoring system.

*B. Linearity*

The linear behaviour of the monitoring system was also studied so as to ensure its identical performance at different temperatures, specifically in the range over which the human body temperature can vary. Such characteristic also enables one to interpolate data points locate in between to adjacent measurements if higher resolution is required. To investigate the linearity behaviour, a controlled heater whose output tem-perature can be changed from 30 C to 40 C was used. Also, an accurate mercury thermometer acting as a reference was placed in a reasonably close distance to the heater beside the wireless thermometer. Figure 10 shows the obtained results from the linearity test. As the results reveals, the monitoring system shows a linear behaviour over the temperature range varying from 30 to 40 C. The mean squared error (MSE) of the measurement results over a range of 10 C, with respect to the reference points, was approximately 0.357 degrees. However, the deviation of the results obtained from the thermometer for a single measurement is limited by the accuracy of the temperature sensor (i.e., 0.4 C).

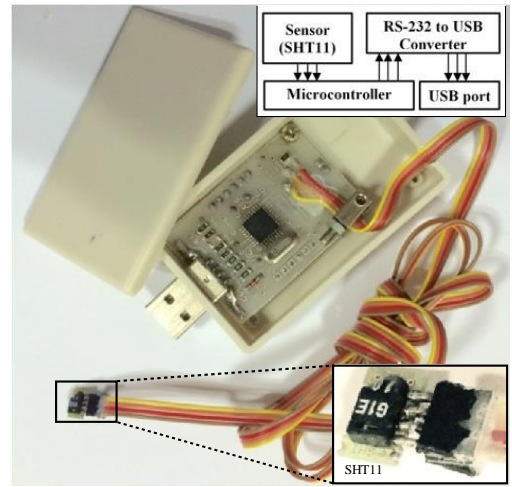

Fig. 8. A picture of the developed wired thermometer.

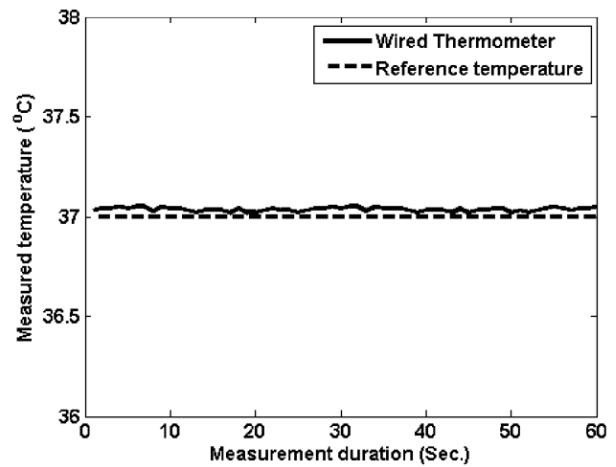

Fig. 9. A volunteer's human body temperature obtained by using the wired thermometer.

*C. Response Time*

Besides the requirements such as linearity and stability, the developed monitoring system needs to be fast enough in order to detect two different types of variation in temperature namely, (1) high body temperature, and (2) rapid increase of patients' body temperature. Therefore, the monitoring system should fulfill a minimum measurement speed (desired response time) so as to be sensitive enough to even small variation in the human body temperature. To estimate the system response time, the controlled heater used in section III-B was set to produce different output temperature at different speed. The heater temperature was increased at the speed of 0.97 degrees per second for 15 seconds and then at the speed of 4.3 degrees per second for next 13 seconds. As it can be seen in Fig. 11, the wireless and mercury thermometers follow each other where the rate of temperature variations is low, whereas this is not the case when the variation rate is high. Finally, the heater output temperature was kept constant after almost 30 seconds in order to show the superior performance of the monitoring system compared to the mercury thermometer which is extensively used in medicine.

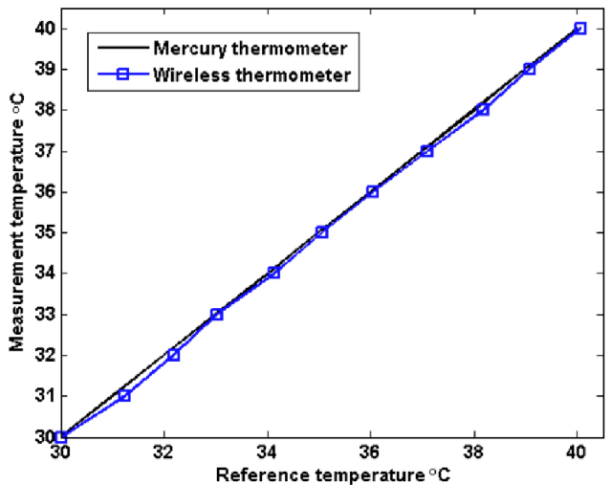

Fig. 10. The linearity test result of the developed temperature monitoring system.

*D. Agility Test*

One of the perspective application of the developed monitoring system is to be used as a portable wireless thermometer. In this application, it is of paramount importance to investigate the agility of the the system in measuring patients' body temperature. To investigate the agile behaviour of the system, a wireless thermometer was set to measure a patient's body temperature for a contact duration of $t_d$ seconds and the latest temperature measurement ($t = t_d$) was buffered and then recorded. After the completion of a round of measurement, the thermometer was held away in rest until the room temperature of 25 C is measured. Next, a series of measurements for different $t_d$ values ranging from 2 up to 18 seconds with a step of 2 seconds was performed. Figure 12 shows the results obtained from the agility test experiment. It can be deduced that the systems can measure a patient's body temperature with approximately $t_d$=12 second which is almost 10 times faster than using a mercury thermometer [17].

## IV. INDOOR PERFORMANCE TEST OF THE TEMPERATURE MONITORING SYSTEM

As the developed temperature monitoring system is intended to be used in medical center (i.e., ambient environment),

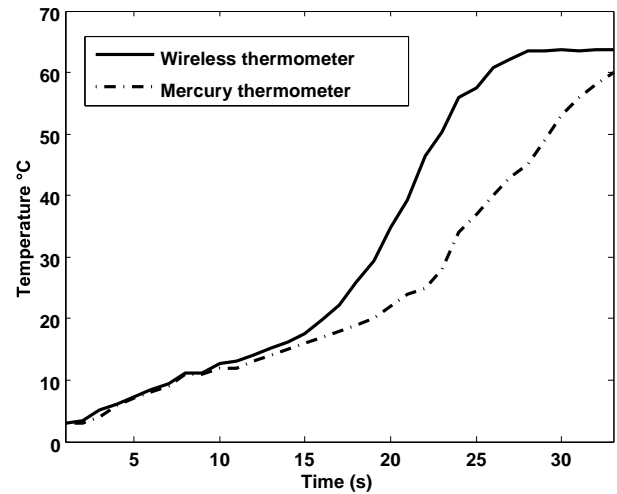

Fig. 11. Response time of the temperature monitoring system.

phenomena such as ambient multiple reflections, multi-path communication in addition to fading can be severely influential on the connectivity between the thermometers with the central node which is connected to a PC for collecting the measurement results. Such phenomena can result in loss of information and consequently an increased delay associated with the measurements. To investigate well performance of the monitoring system for indoor use, the system was tested in different scenarios that closely replicate practical conditions.
To this end, a thermometer was set to measure the body temperature of a volunteer in a room with (S1) and without (S2) furniture. Also, a set of measurement was performed while the measuring thermometer was continuously moved away from the central (S3). Lastly, a line-of-sight scenario (S4) was tested. The distance between the pair of thermometer and central node was varied from 10 to 50 meters with a step of 10 meters. To quantify the performance of the monitoring system, the connectivity term was defined which is calculated based on the consistency and accuracy of measurements. For each scenario a 60-second long measurement was performed and the connectivity values were obtained (Fig. 13). As the results reveal, the connectivity between the communicating pair is significant for distances below 30 meters. Also, the connectivity values for a separation of 50 meters between the thermometer and central node is shown in the inset of Fig. 13. In all none line-of-sight cases where obstacles present and prevent direct communication, the desired performance (i.e., high connectivity) of the monitoring system degrades, that results in a limited coverage area.

## V. CONCLUSIONS

In this paper, a comprehensive description of the design and implementation of a real-time temperature monitoring system was presented. The developed system consists of a number of wireless thermometers that each thermometer incorporates a wireless transceiver unit operating at 2.4 GHz, an accurate digital temperature sensor as well as a microcontroller. Additionally, thermometers were provided with unique identification codes to enable the system to distinguish the relevant measurements. In order to coordinate the thermometers communication and also to collect the measurements reference (i.e., Mercury thermometer). The total MSE of the system is approximately 0.357 C. Also, the performance of the monitoring system was evaluated in an ambient environment (i.e., indoor) and the obtained results revealed that the wireless thermometers can be connected with the central node from a distance of nearly 30 meters without encountering disruption to communication. However, measurements at farther distance (>30 meters) is possible in cases where a direct communication between thermometers and the central node can be maintained.

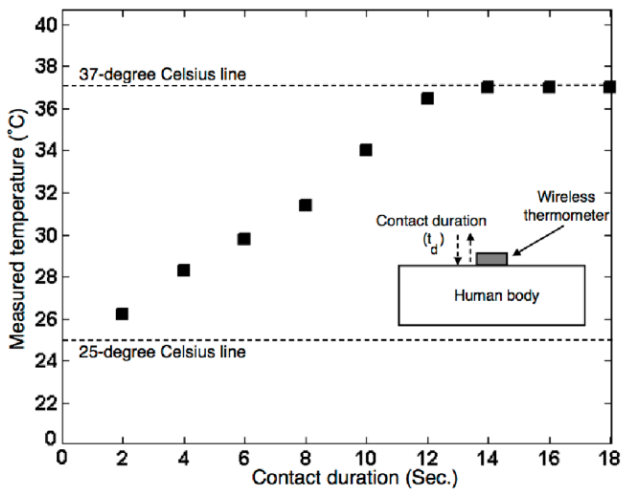

Fig. 12. Agility test results of the monitoring system obtained from measurement.

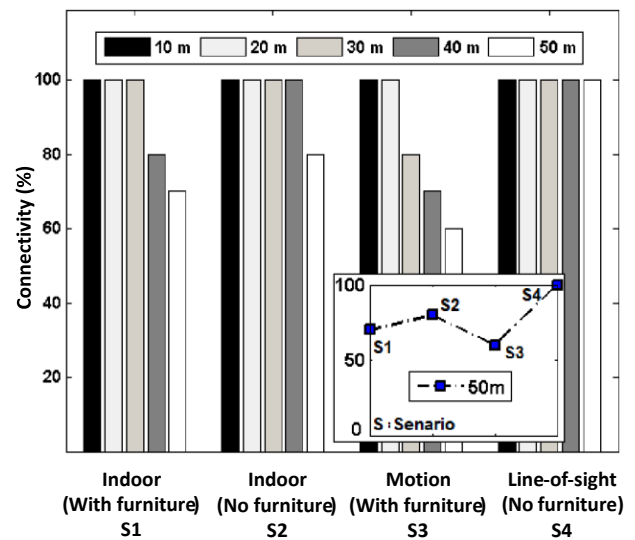

Fig. 13. Calculated connectivity between the measuring thermometer and the central node.

data, a central node was also conceived and implemented. The temperature monitoring system was evaluated in different scenarios for its response time, stability and linearity as well as its agility. The system evaluation results showed a good stability condition with a fluctuation level (averaged) less than 0.25 C. The mean squared error (MSE) was also calculated over a range of 10 C by comparing the obtained results with a